\documentclass{aa}
\usepackage[varg]{txfonts}
\usepackage{graphicx}
\usepackage{txfonts}
\usepackage{natbib}
\usepackage{verbatim}
\usepackage[pdftex]{color}
\usepackage[dvipsnames]{xcolor}
\usepackage{hyperref}
\usepackage{ulem}
\usepackage{subfigure}
\hypersetup{
    colorlinks = true,
    citecolor ={blue}
}
\usepackage{mathrsfs}
\bibpunct{(}{)}{;}{a}{}{,} 

\usepackage{booktabs}
\definecolor{ultramarine}{rgb}{0.07, 0.1, 0.6} 
\definecolor{myblue}{rgb}{0.07, 0.2, 0.6} 
\definecolor{dopal}{rgb}{.70, .25, .05}

\begin{document}

\title{Exsolution process in white dwarf stars 
}

\author{Maria Camisassa\inst{1}\thanks{E-mail: maria.camisassa@upc.edu}, 
Denis A. Baiko \inst{2}, Santiago Torres\inst{1,3}, Alberto Rebassa-Mansergas\inst{1,3}}
\institute{Departament de F\'\i sica, 
           Universitat Polit\`ecnica de Catalunya, 
           c/Esteve Terrades 5, 
           08860 Castelldefels, 
           Spain
           \and
 Ioffe Institute, Politekhnicheskaya 26, 194021 Saint Petersburg, Russia           
           \and
           Institute for Space Studies of Catalonia, 
           c/Gran Capit\`a 2--4, 
           Edif. Nexus 104, 
           08034 Barcelona, 
           Spain}
\date{Received ; accepted }
\abstract{White dwarf stars are considered cosmic laboratories to study the physics of dense plasma. Furthermore, the use of white dwarf stars as cosmic clocks to date stellar populations and main sequence companions demands an appropriate understanding of the white dwarf physics in order to provide precise ages for these stars.
}   
{
We aim at studying exsolution in the interior of white dwarf stars, a process in which a crystallized ionic binary mixture separates into two solid solutions with different fractions of the constituents. Depending on the parent solid mixture composition, this process can release or absorb heat, thus leading to a delay or a speed-up of white dwarf cooling.
}
{
Relying on accurate phase diagrams for exsolution, we have modeled this process in hydrogen(H)-rich white dwarfs with both carbon-oxygen (CO) and oxygen-neon (ONe) core composition, with masses ranging from 0.53 to 1.29 $M_\odot$ and from 1.10 to 1.29 $M_\odot$, respectively.}
{
Exsolution is a slow process that takes place at low luminosities ($\log(L/L_\sun)\lesssim -2.75$) and effective temperatures ($T_{\rm eff} \lesssim 18\,000 K$) in white dwarfs. We find that exsolution begins at brighter luminosities in CO than in ONe white dwarfs of the same mass. Massive white dwarfs undergo exsolution at brighter luminosities than their lower-mass counterparts. The net effect of exsolution on the white dwarf cooling times depends on the stellar mass and the exact chemical profile. For standard core chemical profiles and preferred assumptions regarding miscibility gap microphysics, the cooling delay can be as large as $\sim 0.35$ Gyrs at $\log(L/L_\sun)\sim -5$. We have neglected a chemical redistribution possibly associated with this process, which could lead to a further cooling delay. Although the chemical redistribution is known to accompany exsolution in binary solid mixtures on Earth, given the solid state of the matter, it is hard to model reliably, and its effect may be postponed till really low luminosities.}
{
Exsolution has a marginal effect on the white dwarf cooling times and, accordingly, we find no white dwarf branches on the {\it Gaia} color magnitude diagram associated with it. {  However, exsolution in massive white dwarfs can alter the faint end of the white dwarf luminosity function, thus impacting white dwarf cosmochronology.} }

\titlerunning{Exsolution process in white dwarfs}
\authorrunning{Camisassa et al.}

\maketitle

\section{Introduction}
\label{introduction}

White dwarf stars are the most common stellar remnants, as more than 95\% of main sequence stars end their lives as white dwarfs. These old and numerous objects thus provide valuable information on the stellar evolution theory, the final states of planetary systems, the structure and evolution of our Galaxy, and the properties of stellar matter under extreme conditions ---  see  for  instance, the  reviews  of
\cite{2008PASP..120.1043F},       \cite{2008ARA&A..46..157W},      
\cite{2010A&ARv..18..471A}, \cite{2016NewAR..72....1G}, \cite{2019A&ARv..27....7C}, and \cite{2022FrASS...9....6I}.
Indeed, white dwarfs can be used to infer the star formation rate, the initial-mass-function, the initial-to-final-mass relation, the age-velocity dispersion relation, and the chemical evolution in the solar neighborhood \citep[e.g.][]{2021MNRAS.505.3165R,2022A&A...658A..22R}. 

Furthermore, nuclear reactions have ceased in white dwarf cores and, therefore, these stars undergo a slow cooling process that lasts for several billion years (Gyrs). 
 The use of accurate evolutionary sequences, based on  a proper understanding of the physics involved in the cooling of white dwarfs, allows one to determine the age of individual white dwarfs with  high precision. Thus, these objects have become reliable cosmochronometers to date stellar populations and main sequence companions. In particular, the white dwarf population in open and globular clusters, can provide age estimates independent of the ages determined using the traditional main sequence turn-off method. 
For instance, \cite{2010A&A...513A..50B}, \cite{2010ApJ...708L..32B,2015MNRAS.448.1779B}, and \cite{2023MNRAS.524..108G} studied the white dwarf members to determine the age of the young open clusters, M 67, NGC 2158, NGC 6819,  and M 37, respectively, and  \cite{2002ApJ...574L.155H,2009ApJ...697..965B,2013ApJ...769L..32B,Torres2015}  employed this method to date the globular clusters M 4, $\omega$ Centauri, and NGC 6397. Furthermore, \cite{2010Natur.465..194G} determined the age of the old, metal-rich, open cluster NGC 6791 using the white dwarf cooling sequence, thus solving a long-standing problem of the discrepancy between the main-sequence and white dwarf ages in that cluster.

In order to exploit the white dwarf stars in the Galaxy as powerful cosmic clocks, from the theoretical point of view, we need accurate white dwarf evolutionary models which provide precise stellar ages. To fill this theoretical gap, during the last decades, a new generation of precise white dwarf evolutionary models has been calculated using different stellar evolutionary codes \citep[see, for instance,][]{Camisassa2016,2020ApJ...901...93B,2022MNRAS.509.5197S,2023ApJ...950..115B}. According to \citet{2013A&A...555A..96S}, the uncertainties in the white dwarf cooling times arising from different numerical implementations of the stellar evolution equations nowadays are below 2\%, which is lower than the differences arising from the uncertainties in the chemical stratification. To reduce these latter uncertainties, many white dwarf models employ initial chemical profiles which result from the complete treatment of the progenitor evolution starting on the Zero Age Main Sequence (ZAMS). Then, during the white dwarf phase, the  evolution of the chemical profile due to the action of chemical diffusion and convective mixing is followed \citep{Camisassa2016, Camisassa2017,2019A&A...625A..87C, 2015A&A...576A...9A,ALTUMCO2021}. Furthermore, this new generation of white dwarf models includes all the relevant sources and sinks of energy that govern their evolution. Among these sources, it is important to mention the energy released by sedimentation of neutron-rich species (such as $^{22}$Ne) and the energy released during the process of core crystallization, which releases latent heat and gravitational energy due to phase separation of the main chemical components \citep[see][for details of the implementation of these energy sources]{2023ApJ...950..115B,2022MNRAS.511.5198C}. 
 
Despite the large efforts in the modeling, recent observations have questioned the capability of these models to accurately provide precise white dwarf ages, particularly for the most massive white dwarfs. The {\it Gaia} space mission has revealed the existence of a pile-up in the color-magnitude diagram, named the Q branch, which coincides with the crystallization process \citep{2018A&A...616A...1G,2019Natur.565..202T}. However, recent studies have shown that a fraction of the ultra-massive white dwarfs ($M_{\rm WD} \gtrsim 1.05 M_\odot$) should have a strong delay ($\sim 8$ Gyrs) of their cooling times when over-passing the Q branch. This is not compatible with the cooling delays predicted by models that include the energy released by crystallization, even when both the latent heat and the energy released by phase separation are taken into account. This phenomenon is referred to as ``the cooling anomaly of ultra-massive white dwarfs'' and it is reflected both in their {\it Gaia} kinematic \citep{Cheng2019} and photometric properties \citep{2021A&A...649L...7C}. Although   for massive  ($0.9 M_\odot \lesssim M_{\rm WD} \lesssim 1.1 M_\odot$) white dwarfs, \cite{2019Natur.565..202T} claimed that the energy released by crystallization could account for pile-up in the Q branch, their simplified treatment of the energy released during crystallization leaves an open possibility that a ``cooling anomaly'', probably less prominent, could also take place in these stars. 

On the other hand, \cite{Kilic2020} 
found on the SDSS footprint a well defined ultracool IR-faint branch in the white dwarf color-magnitude diagram, also reported by \cite{2022RNAAS...6...36S} as a faint blue white dwarf branch on the {\it Gaia} color magnitude diagram. \cite{2022ApJ...934...36B} analyzed these IR-faint white dwarfs using accurate mixed-H/He atmospheres and obtained significantly
higher effective temperatures and larger stellar masses than previously estimated. Still, these authors found a group of ultra-massive white dwarfs in the Debye cooling regime, a rapidly cooling phase where, due to ion quantum effects, the ion contribution to the heat capacity of crystallized matter switches from nearly temperature-independent regime to dropping as $T^3$ \citep[see][]{2007A&A...465..249A}. As mentioned by \cite{2022ApJ...934...36B}, it is surprising to find a large number of massive IR-faint white dwarfs caught in this rapidly evolving phase. Therefore, we can speculate on the possibility that an unknown energy source may be causing a delay of the cooling times of these stars.

Several studies have proposed alternative energy sources that could alter the white dwarf evolutionary timescales causing a ``cooling anomaly'' in ultra-massive white dwarfs \citep[see, for instance,][]{2020ApJ...902...93B,2020PhRvD.102h3031H,2021ApJ...919L..12C,2023ApJ...946...78C}. Among these studies, the most promising solutions invoke $^{22}$Ne distillation \citep{2021ApJ...911L...5B} and $^{22}$Ne sedimentation processes \citep{2021A&A...649L...7C} occurring in ultra-massive white dwarfs with CO cores. All attempts to reproduce necessary cooling delay in ultra-massive white dwarfs with ONe cores have failed, and this is particularly interesting because, to date, there is no clear evidence whether ultra-massive white dwarfs harbour CO or ONe cores \citep[see e.g.][for a thorough discussion]{2022MNRAS.516L...1C}.

In two recent papers, \cite{2022MNRAS.517.3962B,2023MNRAS.522L..26B} studied the phase diagrams for crystallization of fully ionized binary ionic mixtures typical of white dwarf interior and found a lower-temperature solid-solid structure never reported in the astrophysical literature before. Instead of a single stable solid solution, they obtained an extensive unstable domain, where a mechanical mixture of two solids is thermodynamically preferable. Such regions are called {\it miscibility gaps} and are well-known in terrestrial materials \citep[e.g.][]{Gordon1968}. The miscibility gaps in azeotropic fully ionized binary mixtures (such as CO or ONe) were demonstrated to arise naturally (with decrease of the charge ratio) from solid-solid regions of eutectic phase diagrams characteristic of mixtures of more different constituents (e.g. CNe, CMg, or OSi). 

After crystallization takes place, white dwarf matter cools down as a single stable solid solution until the temperature reaches the miscibility gap. Then, {\it exsolution} starts taking place, whereby a new solid solution nucleates from the original one. The composition of the new solution is determined by the phase diagram. With further temperature decrease, fraction of the parent solution decreases, whereas fraction of the new one increases. By the end of the process (at $T=0$), the original solid is separated in two solid phases, which, for fully ionized binary ionic mixtures under consideration, have pure ionic compositions. 

The exsolution process can either release or absorb energy, depending on the initial composition of the solid, and thus it can slow down or accelerate white dwarf cooling. In this paper, we study the exsolution process taking place in the interior of both CO-core and ONe-core white dwarfs. To that end, we have incorporated the exsolution process consistently into the {\tt LPCODE} stellar evolutionary code \citep[][]{2005A&A...435..631A}. The energy released by exsolution has been coupled to the luminosity equation on each time step. Thus, we have calculated, for the first time, white dwarf evolutionary models which include the exsolution process and provide realistic cooling times for this process. These models cover a mass range from 0.53 to 1.29 $M_\odot$.
Furthermore, we analyze actual observational samples to identify possible white dwarf candidates in which the exsolution process should be occurring.

To complete the Introduction, let us note that with further improvements of observational capabilities (e.g. recent commissioning of JWST), one can anticipate a steady flow of new and/or more detailed information on the structure of the white dwarf color-magnitude diagram at low luminosities. This makes theoretical research and numerical simulations of various physical processes at later stages of white dwarf evolution even more actual.

\section{Methods}

\subsection{Input physics}
\label{input}

The white dwarf evolutionary models presented in this study were computed using the stellar evolutionary code {\tt LPCODE} \citep[see][for details]{2005A&A...435..631A}. This code has undergone rigorous testing against other evolutionary codes, covering different phases of stellar evolution, including the red giant phase \citep{2020A&A...635A.164S,2020A&A...635A.165C}
and the white dwarf phase \citep{2013A&A...555A..96S}. Furthermore, {\tt LPCODE} has been extensively applied in the study of diverse aspects of low-mass stars \citep[see][and references therein]{2010ApJ...717..897A,2010ApJ...717..183R,2016A&A...588A..25M,Camisassa2017}.

Here, we provide a brief overview of the key input physics utilized in {\tt LPCODE} for modeling the white dwarf regime. For the low-density regime, we adopt the equation of state from \cite{1979A&A....72..134M}, while for the high-density regime, we rely on the equation of state developed by \cite{1994ApJ...434..641S}. Radiative opacities are taken from the OPAL project \citep{1996ApJ...464..943I}, except for the low-temperature regime where {\tt LPCODE} incorporates molecular radiative opacities that account for varying carbon-to-oxygen ratios \citep{2005ApJ...623..585F}. Conductive opacities are taken from \cite{2007ApJ...661.1094C}. To account for neutrino emission rates, including pair, photo, and bremsstrahlung processes, we adopt the rates from \cite{1996ApJS..102..411I}, whereas for plasma processes, we follow the prescription from \cite{1994ApJ...425..222H}. External boundary conditions for our evolving models are determined using non-gray model atmospheres sourced from \cite{2012A&A...546A.119R}. {\tt LPCODE} also considers variations in chemical abundances arising from element diffusion, including gravitational settling, chemical diffusion, and thermal diffusion processes. To model the crystallization process, we employed the phase diagram of \cite{2010PhRvL.104w1101H} for CO mixture and a phase diagram for ONe mixture based on the approach of \cite{2010PhRvE..81c6107M}. To facilitate a comparison with previous work, we do not include the latent heat of crystallization values calculated by \cite{2023MNRAS.522L..26B}, but use a fixed value of 0.77 $k_{\rm B} T$ per ion instead. We do not expect this approximation to qualitatively affect our conclusions.

\subsection{Initial white dwarf models}

We have modeled H-rich CO-core white dwarfs with masses 0.53, 0.58, 0.66, 0.83, 0.95, 1.1, 1.16, 1.23,  and  1.29 $M_\sun$ and H-rich ONe-core ultra-massive white dwarfs with masses 1.1, 1.16, 1.23,  and  1.29 $M_\sun$. For each mass, we have calculated four evolutionary sequences, one in which both processes, crystallization and exsolution, are taken into account, two in which only one process is considered, and the last one in which both processes are disregarded\footnote{  In sequences which disregard crystallization, we switch off latent heat release and phase separation including element redistribution and gravitational heat release. All the other thermodynamic properties of matter below crystallization temperature remain such as they are in the crystallized phase.}. A total of 52 evolutionary sequences has been calculated and we have neglected $^{22}$Ne sedimentation process in all of them. 

The initial chemical profiles and the thermo-mechanic structure of the CO-core models with 0.53, 0.58, 0.66, and 0.83 $M_\sun$, are taken from \cite{2016A&A...588A..25M}. They are the result of the full progenitor evolution from the ZAMS, all the way to the white dwarf phase. For the CO-core 0.95 $M_\sun$ evolutionary models, we have re-scaled the structure of the 0.83 $M_\sun$ model. The ultra-massive CO-core white dwarf models, i.e. models with 1.1, 1.16, 1.23,  and  1.29 $M_\sun$, employ the initial chemical profile that results from the full progenitor evolution of the model stars computed in \cite{ALTUMCO2021} with the Monash-Mount Stromlo code {\tt MONSTAR} \citep{wood1987, frost1996, campbell2008,gilpons2013, gilpons2018}. In this case, the mass-loss rates of
\cite{vassiliadis1993}  were  reduced along the early Asymptotic Giant Branch (AGB) and the 
thermally-pulsing AGB \citep[see][for details of these models]{2022MNRAS.511.5198C}. The initial chemical profiles in the ONe-core models are the same as in \cite{2019A&A...625A..87C}, which result from the progenitor evolution calculated through the Super Asymptotic Giant Branch in \cite{2010A&A...512A..10S}. Note that each initial white dwarf model has a different chemical profile, thus leading to different energy releases from the exsolution process.

In our initial chemical profiles, there are some Rayleigh-Taylor unstable regions, where the mean molecular weight is inverted. These regions are expected to rapidly mix and, therefore, we have performed this mixing at the beginning of the white dwarf phase assuming it to be instantaneous. During the white dwarf evolution, the chemical abundances change due to element diffusion. Gravitational settling purifies the envelopes of our models, leaving  pure-H envelopes, whereas chemical and thermal diffusion smoothes the chemical interfaces. Finally, crystallization process also alters the chemical profile as a result of the phase separation.

\subsection{The exsolution process}
\label{exsolution}

\begin{figure}
        \centering
        \includegraphics[width=1.\columnwidth]{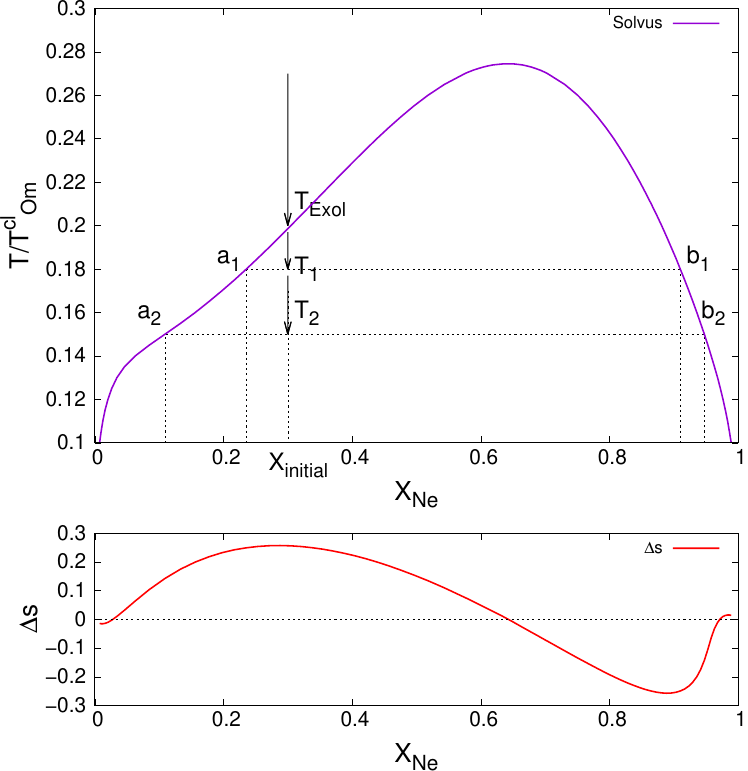}
        \caption{{\it Upper panel:} Phase diagram for the exsolution process in a binary ONe mixture from \cite{2023MNRAS.522L..26B}. The purple line is the solvus and the region below it is the miscibility gap. A solid solution with an initial Ne numerical abundance $X_{\rm initial}$ will start the exsolution at a temperature $T_{\rm Exol}$ determined by the solvus. By the time the temperature drops to the value $T_1$, the solid will be decomposed into two mixtures, one depleted in Ne, with a Ne abundance $X_{\rm Ne}=a_1$, and the other enriched in Ne, with a Ne abundance $X_{\rm Ne}=b_1$. By the time the temperature drops to the value  $T_{2}$, the two solids will have respective abundances $a_2$ and $b_2$. The exsolution process is completed at temperature $T=0$, where the solid is separated into two components, pure O and pure Ne. {\it Lower panel:} Excess entropy of exsolution as a function of Ne numerical abundance in a solid, undergoing decomposition \citep[taken from][]{2023MNRAS.522L..26B}.
        } 
        \label{PD}
\end{figure}

After crystallization, the dense solid matter in the white dwarf interior will undergo the exsolution process, in which the two main components of matter gradually separate. We have modeled this process, following the phase diagrams for binary mixtures of CO and ONe from \citet[Figure 1]{2023MNRAS.522L..26B}. The phase diagram for the exsolution process in an ONe mixture is reproduced in the upper panel of Figure \ref{PD}. We shall omit the CO phase diagram description because it exhibits a similar behaviour. The $x$-axis in Figure \ref{PD} is the Ne numerical abundance and the $y$-axis in the upper panel of this figure is the temperature in units of the crystallization temperature of
the lighter-ion (i.e. oxygen) plasma, $T^{\rm cl}_{\rm Om}=Z_{\rm O}^{5/3} e^2/(k_{\rm B} a_{\rm e} \Gamma_{\rm m})$. In this case, the atomic number is $Z_{\rm O}=8$, $k_{\rm B}$ is the Boltzmann constant, $e$ is the electron charge, $a_{\rm e}=(4\pi n_{\rm e}/3)^{-1/3}$ is the mean electron spacing, $n_{\rm e}$ is the electron density, and the Coulomb coupling parameter $\Gamma_{\rm m}=175.6$.

The temperature of crystallized matter keeps dropping until it reaches the {\it solvus}, which is the purple line in the upper panel of Figure \ref{PD} that delimits the miscibility gap. Any solution inside the miscibility gap is thermodynamically unstable. Thus, the mixture will evolve through the solvus, separating into two mixtures with different O-Ne proportions, determined by the lever rule. For instance, an ONe solid mixture with initial Ne numerical abundance $X_{\rm initial}$ will begin the exsolution at a temperature $T_{\rm Exol}$, determined by the solvus. When the temperature decreases to a value $T_1$, the mixture is going to be divided into two mixtures with different O-Ne proportions, one on the left side of the miscibility gap with Ne numerical abundance $a_1$ and the other on the right side of the gap with Ne numerical abundance $b_1$. When the temperature drops to a value $T_2$, the Ne-depleted solution will have Ne abundance $a_2$ and the Ne-enriched solution will have Ne abundance $b_2$. The exsolution will continue till $T=0$, where O and Ne will be completely separated with $a_{\rm final}=0$ and $b_{\rm final}=1$.

Let us define two more quantities: $N_1$, which is the number of ions on the left side of the gap (i.e. in the Ne-depleted solution) and $M_1$, which is the number of ions on the right side of the gap (i.e. in the Ne-enriched solution). The condition $N_1+M_1=N_{\rm TOT}$, where $N_{\rm TOT}$ is the total number of ions in the initial mixture, must be fulfilled at all times. 
The values of $N_i$ and $M_i$ at a timestep $i$ are determined by:
\begin{eqnarray}
N_i/N_{\rm TOT}&=&(b_i-X_{\rm initial})/(b_i-a_i)~, \\
M_i/N_{\rm TOT}&=&(X_{\rm initial}-a_i)/(b_i-a_i)~. 
\end{eqnarray}

The white dwarf core is not a binary mixture. For the case of CO-core white dwarfs, the total mass fraction of C plus O in the core is $\gtrsim 95\%$, whereas for ONe-core white dwarf, the total mass fraction of O plus Ne is $\gtrsim 85\%$. We have neglected exsolution in the white dwarf envelope, where these fractions monotonically decrease  as a function of the stellar radius (or inner mass fraction). 
For CO-core white dwarfs, having the O abundance $X_{\rm O}$, we consider the carbon abundance to be  $X_{\rm C}=1 - X_{\rm O}$. That is, for modeling exsolution, we consider all other elements that are neither C nor O as C. Similarly, for ONe-core white dwarfs, having the Ne abundance $X_{\rm Ne}$, we consider the O abundance as $X_{\rm O}=1 - X_{\rm Ne}$.

The exsolution is a slow process that lasts for billions of years (formally speaking, it lasts endlessly till $T=0$) and can release or absorb heat. In each evolutionary time-step, we have added this {\it exsolution heat} term to the luminosity equation in the stellar evolutionary code {\tt LPCODE} \citep[equation 1 in][]{2022MNRAS.511.5198C}. The exsolution heat has been taken from the entropy change calculated by \citet[Figure 1c]{2023MNRAS.522L..26B}. This entropy excess is shown by a red line in the lower panel of Figure \ref{PD} for a binary mixture of O-Ne. In the case that the solid starts the exsolution on the left side of the miscibility gap, the heat on the timestep 2 is dictated by the change from $M_1$ to $M_2$:
\begin{equation}
l_{\rm Exol}= k_{\rm B} T_1 (M_2-M_1) \Delta s~,
\end{equation}
where $\Delta s$ is the specific entropy change evaluated at $a_1$. In the case that the solid starts the exsolution on the right side of the miscibility gap, the heat at the timestep 2 will be:
\begin{equation}
l_{\rm Exol}= k_{\rm B} T_1 (N_2-N_1) \Delta s~,
\end{equation}
where $\Delta s$ is evaluated at $b_1$. 

Note that, in this latter case, $\Delta s$ is negative and therefore, the process absorbs energy, leading to a cooling acceleration. {  The condition for reaching the solvus on the left (right) side of the miscibility gap is that the Ne abundance $X_{\rm initial}$ is lower (higher) than $\approx 0.64$. Therefore, those regions where $X_{\rm initial}>0.64$ will absorb energy and those regions where $X_{\rm initial}<0.64$ will release energy. In all regions of all our ONe white dwarf models, $X_{\rm initial}<0.64$.  For CO mixtures, the solvus maximum is at the O abundance $\approx 0.66$ and, in some regions of our CO models, $X_{\rm initial}$ can be larger than 0.66, thus leading to the appearance of an energy sink.}

It is possible that a chemical redistribution {  in which the heavier solution sinks towards the center of the star} occurs upon exsolution, despite the fact that the matter is in a solid phase. On Earth, exsolution in the solid phase results in a formation of lamellar structures from microscopic to visible by naked eye sizes. How this translates into a white dwarf is an open question. {  In the present study, we simply neglect any ensuing chemical redistribution  and include only the exsolution heat release due to the difference in entropy}.

There are other theoretical uncertainties in the description of the exsolution process. In this paper, we rely on binary mixture thermodynamics derived from the linear-mixing formalism with corrections to linear-mixing energies reported by \citet{O+93}. In principle, the actual miscibility gaps can be more or less prominent than we have here, if another parametrization of these corrections is used, if formation of an ordered binary crystal plays any role, or if, effectively, the spinodal curve has to be used in place of the solvus \citep[see][for discussion]{2022MNRAS.517.3962B}. The entropy difference would have to be then re-calculated accordingly.

\section{Results}
\label{Results}

\subsection{General features of the cooling tracks}
\label{gen_cooling_track}

\begin{table*}
\begin{center}
 
\caption{Effective temperature and stellar luminosity at crystallization and exsolution onsets for each of the white dwarf models presented in this paper. Also listed are the delays, $\Delta t_{\rm cool}$, in Gyrs (and the relative delays, $\Delta t_{\rm cool}/t_{\rm cool}^{(0)}$, in parenthesis) of white dwarf cooling times at the stellar luminosity $\log(L/L_\odot)=-5$ caused by exsolution alone, crystallization alone, and both processes combined, with respect to the cooling sequence that disregards both processes.}
\begin{tabular}{cccccccc}
\hline
\hline
              &\multicolumn{2}{c}{crystallization} & \multicolumn{2}{c}{exsolution}& 
\multicolumn{3}{c}{Total delay in Gyrs (relative delay) at $\log(L/L_\odot)=-5$} \\
    \cline{6-8}
  $M/M_\odot$  
  & \multicolumn{2}{c}{onset}             & \multicolumn{2}{c}{onset} 
  & exsolution w/o &
crystallization & exsolution + \\
             & $\log{(L/L_\odot)}$ & $T_{\rm eff}$& $\log{(L/L_\odot)}$ & $T_{\rm eff}$   & crystallization & w/o
exsolution & crystallization \\
\hline
\multicolumn{8}{c}{CO-core} \\
\hline
  0.53 &-3.94&5\,138&-4.60&3\,552&0.004 (4$\times 10^{-4}$) &1.45 (0.12) &1.49 (0.12)\\
  0.58 &-3.88&5\,492&-4.45&3\,973&0.1 (6$\times 10^{-3}$)&1.7 (0.11) & 1.78 (0.12)\\
  0.66 &-3.70&6\,361&-4.34&4\,435&0.18 (0.013) &1.87 (0.14) & 2.02 (0.15)\\
  0.83 &-3.24&9\,120&-4.15&5\,413&0.3 (0.02)&1.92 (0.15)&2.2 (0.17) \\
  0.95 &-2.95&11\,522&-3.93&6\,582&0.35 (0.03)&1.68 (0.14) & 2.0 (0.17)\\
  1.10 &-2.51&16\,615&-3.53&9\,277&0.25 (0.03) &1.17 (0.12) & 1.4 (0.15)\\
  1.16 &-2.34&19\,450&-3.33&10\,949&0.25 (0.03)&0.97 (0.11)&1.21 (0.14)\\
  1.23 &-2.06&24\,732&-3.05&13\,914&0.22 (0.03)&0.74 (0.11)&0.97 (0.14)\\
  1.29 &-1.76&32\,316&-2.74&18\,253&0.18 (0.04)&0.54 (0.11) &0.73 (0.15)\\
\hline
\multicolumn{8}{c}{ONe-core} \\
\hline
  1.10 &-2.19&20\,089&-3.91&7\,519&0.12 (0.01)&0.28 (0.03)&0.39 (0.05)\\
  1.16 &-2.01&23\,551&-3.75&8\,749&0.13 (0.02)&0.18 (0.02)&0.31 (0.04)\\
  1.23 &-1.81&28\,785&-3.48&11\,143&0.11 (0.02)&0.003 (6$\times 10^{-4}$) &0.11 (0.02)\\
  1.29 &-1.52&37\,761&-3.16&15\,081&0.09 (0.02)&-0.05 (-0.01)&0.03 (0.01)\\
\hline
\end{tabular}
\end{center}
\label{num_models}
\end{table*}

\begin{figure}
        \centering
        \includegraphics[width=1.\columnwidth]{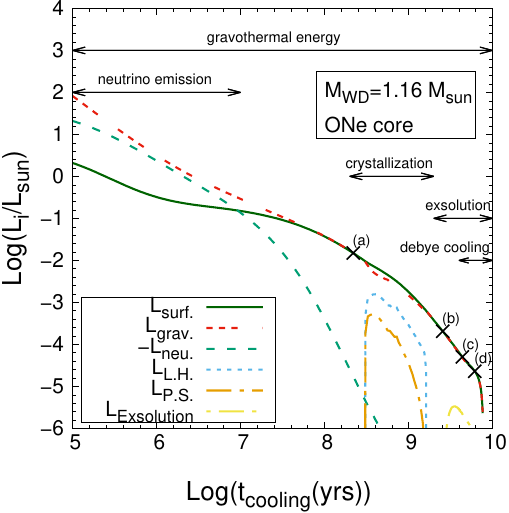}
        \caption{Main contributions to the white dwarf luminosity (solid dark-green line)
  in terms of the cooling age (defined as zero   when the star reaches the maximum effective temperature) for 1.16 $M_\odot$ ONe model including crystallization and exsolution processes. The dashed red line displays the gravothermal energy release, the dashed green line indicates the energy lost by neutrino emission, the dotted blue line is the latent heat of crystallization, the dot-dashed orange line shows the energy released due to the ONe phase separation induced by crystallization, and the dot-dashed yellow line displays the exsolution heat. The arrows indicate the times when these physical processes are acting. The crosses indicate the moments at which we plot the chemical profiles in Figure \ref{perfil}.} 
        \label{leo}
\end{figure}

\begin{figure*}
        \centering
        \includegraphics[width=2.\columnwidth]{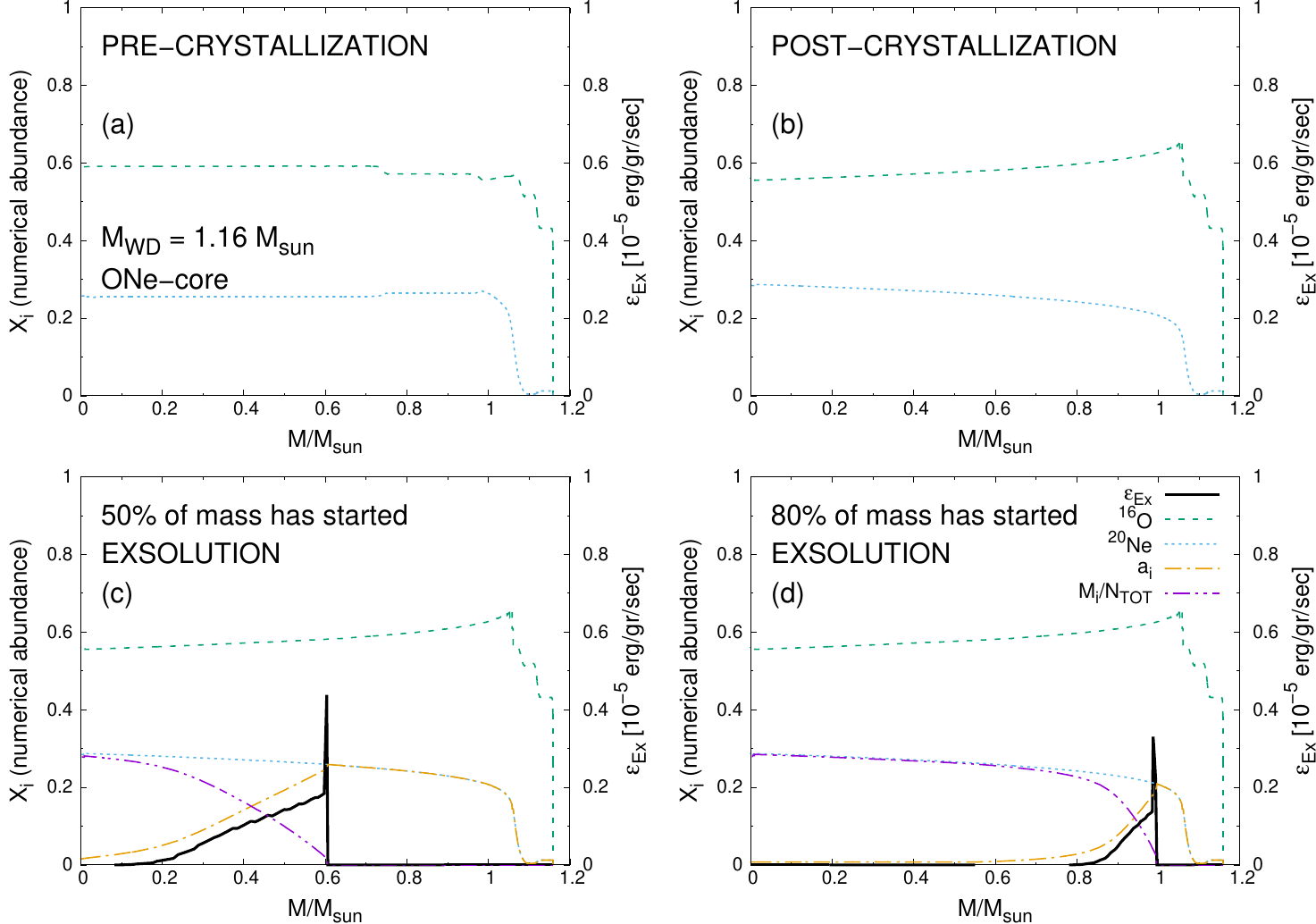}
        \caption{Chemical profiles and exsolution heat release for our 1.16 $M_\odot$ ONe-core model including crystallization and exsolution at different stages of the white dwarf evolution. The dashed green (dotted blue) line displays the $^{16}$O ($^{20}$Ne) numerical abundance, the solid black line depicts the energy released by the exsolution process in units of $10^{-5}$erg/gr/s, the dot-dashed orange line displays the numerical abundance $a_i$ of the Ne-depleted mixture, and the dot-dot-dashed purple line depicts the fraction of ions in the Ne-enriched mixture $M_i/N_{\rm TOT}$ (see Section \ref{exsolution} for details). Panels (a), (b), (c), and (d) show the moments before the crystallization onset ($T_{\rm eff}=25\,700$K), after 95\% of the mass has crystallized ($T_{\rm eff}=8\,900$K), when 50\% of the white dwarf mass has started the exsolution ($T_{\rm eff}=6\,400$K), and when 80\% of the white dwarf mass has started the exsolution ($T_{\rm eff}=5\,300$K), respectively. These epochs are marked in Figure \ref{leo} by crosses.}
        \label{perfil}
\end{figure*}

The main contributions to the white dwarf luminosity as a function of the logarithm of the white dwarf cooling time for our 1.16 $M_\odot$ ONe-core model are shown in Figure \ref{leo}. We define the white dwarf cooling time as the time elapsed since the star reached the maximum effective temperature when entering the white dwarf phase. 
We can see that, through the entire white dwarf phase, the gravothermal energy dominates the white dwarf energetics. Also, at early stages of the white dwarf cooling sequence, the energy lost by neutrino emission is important and can be of the order of the gravothermal energy. 
At $\log{(t_{\rm cooling})}\sim 8.5$, when the logarithm of the stellar luminosity is $\log(L/L_\odot)\sim -2$, the crystallization process starts in the center of the white dwarf, releasing latent heat and gravitational energy due to the ONe phase separation. Note that, in this model, the energy released as latent heat is $\sim 4$ times larger than the energy released by phase separation. The crystallization in this ONe model will last for $\sim 1.7 \times 10^9$ years. 
Finally, when $\log{(t_{\rm cooling})}\sim 9.4$ and the logarithm of the stellar luminosity is $\log(L/L_\odot)\sim -3.75$ (i.e. $\sim 55$ times lower than at crystallization onset), exsolution process starts in the center of this white dwarf model. The release of the major fraction of the exsolution heat will be occurring over $\sim 2.8\times 10^9$ years. The energy released as exsolution heat is $\sim 2$ orders of magnitude lower than the latent heat released by crystallization. Nevertheless, since the stellar luminosity at exsolution is also much lower, the effect of exsolution on the cooling times is just some 30\% smaller than that of crystallization.

In Figure \ref{perfil}, we show the chemical profiles and the heat released by exsolution in our 1.16 $M_\odot$ ONe-core model at 4 different epochs: (a) before crystallization onset, (b) when $\sim 95$\% of the white dwarf mass has crystallized, (c) when $\sim 50\%$ of the white dwarf mass has started the exsolution, and (d) when $\sim 80\%$ of the white dwarf mass has started the exsolution. These epochs are marked in Figure \ref{leo} by crosses.
Before the crystallization onset, the chemical profile of the star is that resulting from the progenitor evolution calculated by \cite{2010A&A...512A..10S}, and rehomogeneized due to Rayleigh-Taylor instabilities \citep[see][]{2019A&A...625A..87C}. Due to the shape of the phase diagram, after crystallization, the inner regions of the core are additionally enriched in Ne (blue dotted line) and depleted in O (green dashed line), whereas the outer regions are enriched in O and depleted in Ne. This is the chemical profile which will dictate the exsolution process (in models without crystallization, the exsolution will commence with the chemical profile pre-crystallization).

In this model, when the temperature reaches the solvus, it does so always on the left side of the miscibility gap, because $X_{\rm Ne}$ is lower than {  0.64} everywhere. Thus, the exsolution heat is always positive. 
In panel (c), the center of the white dwarf ($M_r/M_\sun=0$) has practically completed the exsolution, thus the heat being released is already zero in this region. By contrast, the layers with $M_r/M_\sun \approx 0.6$ are just starting this process and thus the exsolution heat there is at a maximum. The regions where 
$0.05 \lesssim M_r/M_\sun\lesssim 0.6$ are undergoing the exsolution and the regions where $M_r/M_\sun \gtrsim 0.6$ have not started it yet. After $\sim 1.8$ Gyrs, in panel (d), the regions where $M_r/M_\sun\lesssim 0.7$ have practically finished the exsolution, while the regions with $0.7 \lesssim M_r/M_\sun\lesssim 1$ are still undergoing it. 

All other ONe models exhibit a similar behaviour. The solvus is reached on the left side of the miscibility gap in all the regions of the white dwarf and the exsolution heat is always positive. However, in certain regions of some CO-core white dwarf models, the solvus is reached on the right side of the miscibility gap. Such is the case for the inner regions of our CO-core 0.53, 0.58, 0.66, 1.1, 1.16, 1.23, and 1.29 $M_\odot$ models, where the exsolution heat is absorbed.

\subsection{Effect on the white dwarf cooling times}
\label{cooling_times}

\begin{figure*}
        \centering
        \includegraphics[width=2.\columnwidth]{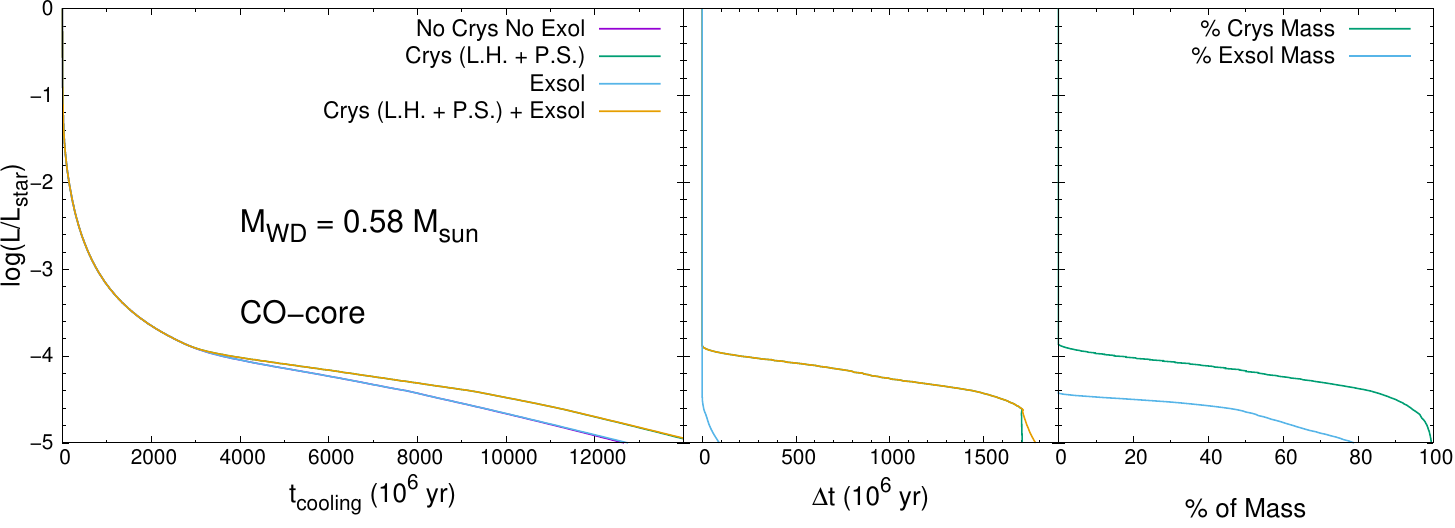}
        \caption{Impact of crystallization (including latent heat and phase separation) and exsolution processes on the cooling times of a 0.58 $M_\odot$ CO-core white dwarf model. The left panel shows the luminosity versus cooling time for the sequence which disregards crystallization and exsolution (purple line), includes crystallization but disregards exsolution (green line), includes exsolution but disregards crystallization (blue line), and includes both processes (orange line). The middle panel shows the cooling delay of the last three sequences with respect to the first one. The right panel shows the percentage of crystallized mass (green line) and the percentage of mass that has started the exsolution (blue line) versus the stellar luminosity.}
        \label{CO058}
\end{figure*}

\begin{figure*}
        \centering
        \includegraphics[width=2.\columnwidth]{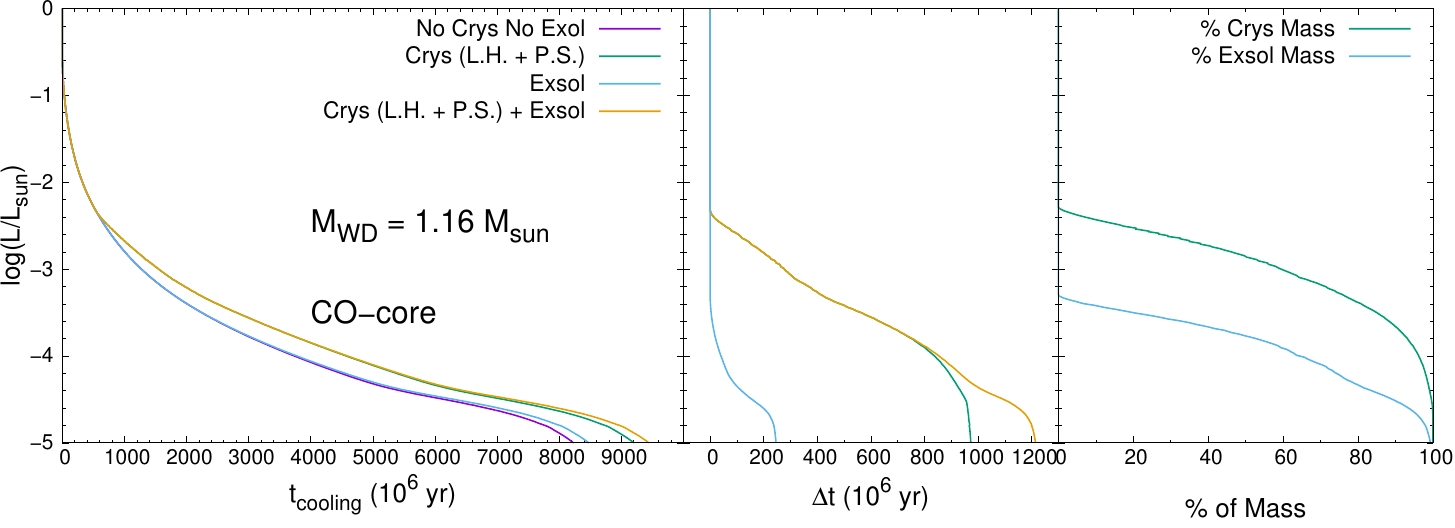}
        \caption{Same as Figure \ref{CO058}, but for the 1.16 $M_\odot$ CO-core white dwarf model.
        }
        \label{CO116}
\end{figure*}

\begin{figure*}
        \centering
        \includegraphics[width=2.\columnwidth]{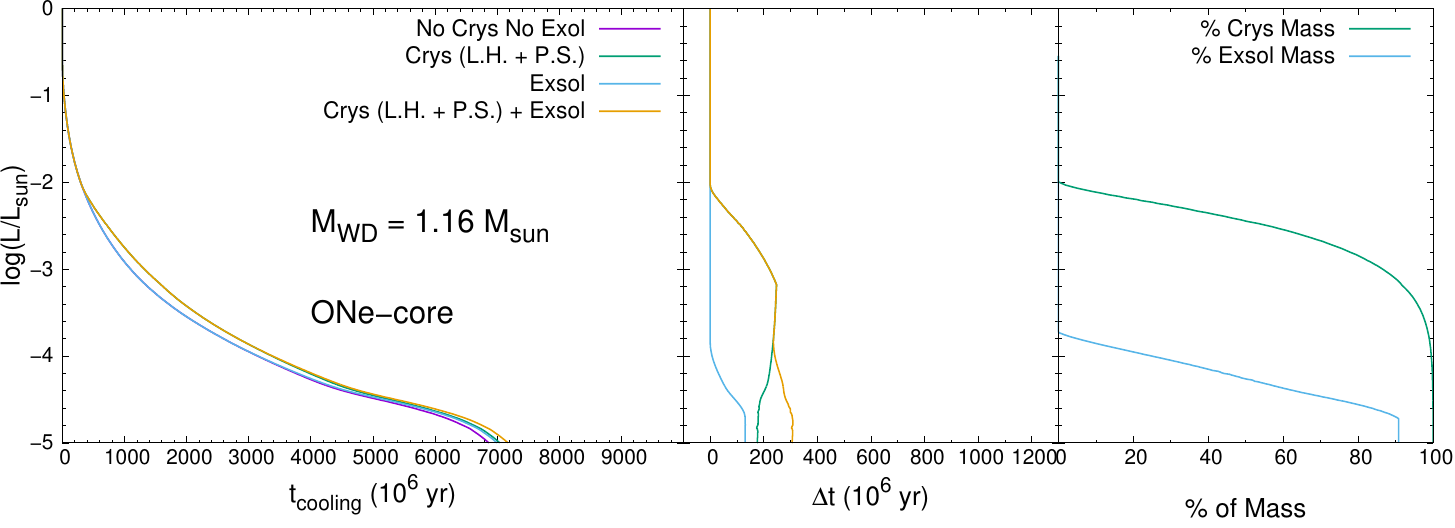}
        \caption{Same as Figure \ref{CO058}, but for the 1.16 $M_\odot$ ONe-core white dwarf model.
        }
        \label{ONe116}
\end{figure*}

The main properties of the 52 white dwarf evolutionary models calculated in our study are summarized in Table \ref{num_models}. For each white dwarf mass, we list the logarithm of the stellar luminosity, $\log{(L/L_\odot)}$, and the effective temperature $T_{\rm eff}(K)$ at the crystallization and exsolution onsets. {  We also list, for each mass, the absolute and relative cooling time delays, $\Delta t_{\rm cool}$ and $\Delta t_{\rm cool}/t_{\rm cool}^{(0)}$, caused by exsolution alone, crystallization alone, and both processes combined, with respect to the cooling sequence that disregards both processes. The relative cooling delays are shown in parenthesis and $t_{\rm cool}^{(0)}$ is the cooling time of the sequence that disregards both processes.}

In concordance with crystallization, exsolution begins at brighter luminosities and higher effective temperatures in more massive white dwarfs (which have higher central densities). On the other hand, although ONe white dwarfs crystallize at brighter luminosities (and effective temperatures) than CO white dwarfs of the same mass, the opposite is true for the exsolution, which occurs at fainter luminosities in ONe white dwarfs. This can be understood from the phase diagrams \citep[see Figure 1a in][]{2023MNRAS.522L..26B}, which predict a higher temperature for exsolution relative to the melting temperature in a CO mixture, than in an ONe mixture. 

The effect of exsolution on the cooling times of selected white dwarf models can be seen in Figures \ref{CO058}, \ref{CO116}, and  \ref{ONe116}. In the left panels of these figures, we show the logarithm of the stellar luminosity, in solar units, versus white dwarf cooling times, for the sequences that disregard crystallization and exsolution processes (purple line), include exsolution but disregard crystallization
(blue line), include crystallization but disregard exsolution (green line), and include both processes (orange line). The cooling delays of these latter three sequences with respect to the sequence that disregards both crystallization and exsolution are plotted in the middle panels. Note that these cooling delays when $\log{(L/L_\odot)}=-5$ are listed in columns 6, 7 and 8 of Table \ref{num_models}, respectively. Finally, in the right panels, we show the percentage of crystallized mass and the percentage of mass that has started the exsolution as a function of the stellar luminosity. 

In Figure \ref{CO058}, we can see that, in the 0.58 $M_\odot$ model, while crystallization (both latent heat and phase separation) produces a delay of $\sim 1.7$ Gyrs at $\log{(L/L_\odot)}=-5$, exsolution produces a marginal delay of $\sim 0.1$ Gyr. Exsolution begins at very faint luminosities in a 0.58 $M_\odot$ star. Indeed, at $\log{(L/L_\odot)}=-5$, although 75\% of the white dwarf mass has started the exsolution, less than 10\% of the mass has released most of its exsolution heat and therefore a lot of this heat is yet to be generated. {  More generally, incomplete heat production by $\log{(L/L_\odot)}=-5$ in low-mass stars is responsible for the initial growth of time delays with mass increase in columns 6 and 7 of Table \ref{num_models}. }
 
More massive white dwarfs undergo crystallization and exsolution at higher luminosities, and such is the case for the 1.16 $M_\odot$ CO model shown in Figure \ref{CO116}. In this model, while the time delay caused by crystallization is $\sim 1$ Gyr at $\log{(L/L_\odot)}=-5$, $\sim 90$\% of the white dwarf mass has already effectively finished the exsolution by this luminosity. This adds another $\sim 0.25$ Gyrs to the time delay.   

It is interesting to note that, in the two CO models above, the oxygen abundance in the liquid phase is less than {  0.66}, but, upon crystallization and redistribution, $X_{\rm O}$ increases above this value in some inner core layers. Thus, these latter layers become a heat sink, which partially offsets the exsolution heat released by the outer layers of the core. The net effect on the white dwarf cooling times is a delay, and the difference between the exsolution delays with and without crystallization illustrates the effect of heat absorption.

{  A peculiar case is represented by the 0.53 $M_\odot$ CO model. In this star, the oxygen abundance slightly exceeds the critical value in the inner $\sim 0.33$ $M_\odot$ even without crystallization, and this whole region becomes a heat sink. The net cooling delay due to exsolution is then mere 4 million years. However, if crystallization is taken into account, only inner $\sim 0.26$ $M_\odot$ operates as a heat sink, noticeably more massive outer region emits heat, and exsolution adds 40 million years to cooling delay due to crystallization. }

Finally, in the 1.16 $M_\odot$ ONe-core model (Figure \ref{ONe116}), the time delays induced by crystallization and exsolution are of the same order at $\log{(L/L_\odot)}=-5$, being $\sim 0.18$ and $\sim 0.13$ Gyrs, respectively. At such low luminosities, $\sim 90$\% of the white dwarf mass has mostly completed the exsolution. Note that, in this model, the percentage of mass that has started the exsolution process (blue line, right panel) increases monotonically as evolution proceeds until it reaches a value of $\sim 90\%$, where it stops growing. The reason for this is that, in the outer 10\% of the white dwarf mass, C and other elements are abundant and we can no longer consider the approximation of a binary mixture of O-Ne. Therefore, we neglected the exsolution in the outer 10\% of the white dwarf mass for this model.

Another relevant feature of this model is that the time delay caused by crystallization starts to decrease as a function of luminosity for $\log{(L/L_\odot)}\lesssim -4$. Such a ``reversal effect'' also occurs in the most massive (1.23 and 1.29 $M_\odot$) ONe-core models, where the time delay caused by crystallization can become zero or negative at very low luminosities. This effect has already been analyzed by \citet{2019A&A...625A..87C} who noticed that, at these late stages, more than 99\% of the white dwarf mass had crystallized and no more energy was released due to crystallization, but the phase separation induced by crystallization had strongly altered both the structure and thermal properties of ultra-massive white dwarfs, impacting their cooling rates. 

Generally speaking, the exact time delay produced by exsolution depends on the white dwarf mass, composition, and chemical profile left by the progenitor evolution and chemical redistribution upon crystallization. For standard core chemical profiles and preferred assumptions regarding miscibility gap microphysics, it can be at most $0.35$ Gyrs at $\log{(L/L_\odot)}\sim -5$.

\subsection{Searching for observational evidence}
 \label{obs}
 
\begin{figure}
        \centering
        \includegraphics[width=1.\columnwidth,
        ,clip]{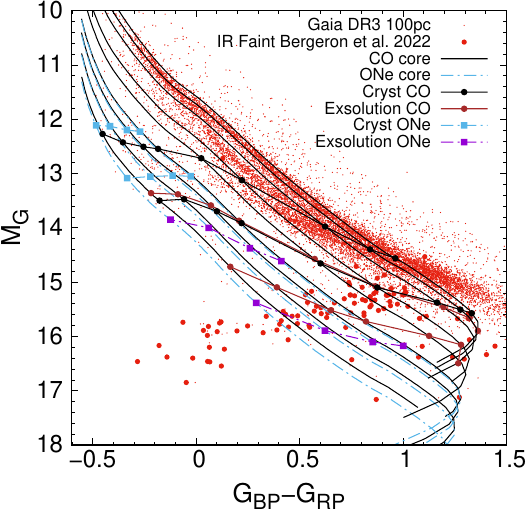}
        \caption{Crystallization and exsolution on {\it Gaia} DR3 color-magnitude diagram. Solid black lines and dot dashed-blue lines are CO and ONe evolutionary sequences, respectively. The masses of the CO (ONe) sequences are, from top to bottom, 0.53, 0.58, 0.66, 0.83, 0.95, 1.1, 1.16, 1.23, and 1.29 (1.1, 1.16, 1.23, and 1.29) $M_\odot$. Filled blue squares (filled black circles) indicate the crystallization onset and the moment when 80\% of the white dwarf mass has crystallized for ONe (CO) white dwarfs. Filled purple squares (filled brown circles) show the exsolution onset and the moment when 80\% of the white dwarf mass has started exsolving for ONe (CO) white dwarfs. The white dwarf sample within 100 pc and the IR-faint white dwarfs from \cite{2022ApJ...934...36B} are plotted by red dots and filled red circles, respectively.
        } 
        \label{gaia}
\end{figure}

\begin{figure}
        \centering
        \includegraphics[width=1.\columnwidth,
        ,clip]{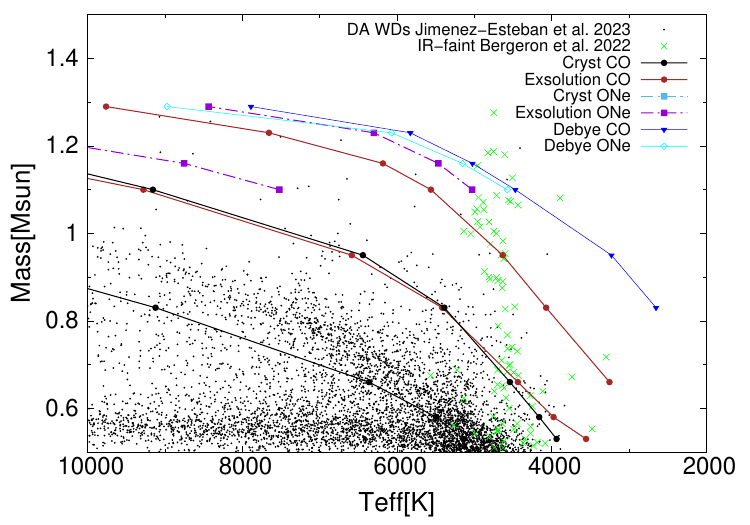}
        \caption{Crystallization and exsolution processes on the mass--effective temperature plane. Solid black lines indicate the crystallization onset and the moment when 80\% of the white dwarf mass is crystallized in our CO models. Solid brown (dot-dashed purple) curves display the exsolution onset and the moment when 80\% of the white dwarf mass has started exsolving in CO (ONe) models. Solid blue and cyan lines indicate the transition between the classic and the quantum Debye regime in dense ionic plasma. Black dots are the DA white dwarfs within 100 pc from the Sun \citep[][]{2023MNRAS.518.5106J}, while the green crosses are the sample of 105  IR-faint white dwarfs \citep[][]{2022ApJ...934...36B}.} 
        \label{teffm}
\end{figure}

In Figure \ref{gaia}, a {\it Gaia} DR3 color-magnitude diagram for the sample of white dwarfs within 100 pc from the Sun is shown along with the IR-faint white dwarfs in the sample from \cite{2022ApJ...934...36B} and our white dwarf cooling tracks. We have employed the pure H atmosphere models of \cite{2010MmSAI..81..921K} to convert our evolutionary models to the {\it Gaia} passbands. The IR-faint white dwarfs from  \cite{2022ApJ...934...36B} are not expected to follow the pure H evolutionary tracks on the {\it Gaia} color-magnitude diagram, as their atmospheric composition is not pure H, but a mixture of H and He instead. Blue squares (filled black circles) indicate the crystallization onset and the moment when 80\% of the mass has crystallized in our ONe (CO) models. Filled purple squares (filled brown circles) delimit the exsolution regions for ONe (CO) white dwarfs. More precisely, they show the exsolution onset (i.e. the moment when the stellar center reaches the solvus) and the moment when 80\% of the mass has started exsolving. 

As can be observed, there is a pile-up of objects between the CO crystallization onset and the 80\% of CO crystallization. This pile-up is the Q branch discussed in the Introduction. We can see that this branch is compatible with the crystallization regions for both CO and ONe models with H atmospheres. 

It is important to remark that exsolution is a slow process. Once it begins in a stellar layer, it will take several evolutionary timescales before the temperature drops down sufficiently (cf.\ Figure \ref{PD}) and the major heat fraction is released (or absorbed). In some of our CO models, the exsolution starts in the center of the star when the outer regions have not yet started crystallizing. 
In Figure \ref{gaia}, we show the moment when matter at the mass coordinate $M_r/M_{\rm WD}=0.8$ reaches the solvus, but the inner regions with $M_r/M_{\rm WD}<0.8$ are still undergoing the exsolution, and the associated heat there is not zero [cf.\ Figure \ref{perfil}(d)]. In a simple inspection, we cannot attribute any structure on the white dwarf color-magnitude diagram to the exsolution process.

In Figure \ref{teffm}, we show the location of crystallization and exsolution processes on the effective temperature vs. mass plane, together with the DA white dwarfs within 100 pc from \cite{2023MNRAS.518.5106J} and the IR-faint white dwarfs from \cite{2022ApJ...934...36B}. The crystallization in ONe white dwarfs occurs at higher effective temperatures than the range of this plot. As mentioned in Section \ref{gen_cooling_track}, the exsolution takes place at higher effective temperatures in CO than in ONe white dwarfs of the same mass, which is in contrast to the crystallization. 

The boundary between classic and quantum Debye regimes at the stellar center is plotted by blue and cyan curves for CO and ONe models, respectively. This boundary is defined as in \cite{2007A&A...465..249A}:
the temperature at the center $T=\theta_{\rm D}/20$, where the Debye temperature is $\theta_{\rm D}=3.48 \times 10^3 \rm \langle Z/A \rangle \rho^{1/2}$ K. We can see that the most massive model with ONe core, with 1.29 $M_\odot$, enters the Debye regime when less than 80\% of its mass has started exsolving. We find that only 7 of the 105 IR-faint white dwarfs analyzed by \cite{2022ApJ...934...36B} are in the Debye cooling phase. 

The IR-faint white dwarfs with masses $M_{\rm WD}/M_\odot \gtrsim 0.66$ should be undergoing the exsolution process. However, the accumulation of IR-faint white dwarfs in a narrow strip around $T_{\rm eff} \sim 4500$ K regardless of the mass is hard to explain by a specific physical process in the core because of the vast range of physical conditions which seem to be involved: from incomplete crystallization in low-mass stars to quantum solids in the ultra-massive ones.  

{  Overall, we find no signatures associated with the exsolution process on both the {\it Gaia} color magnitude diagram and the effective temperature vs. mass plane. This is consistent with the marginal effect of the exsolution process on the white dwarf cooling times. }

\section{Summary and conclusions}
\label{conclusions}

After crystallization, the temperature keeps dropping and the solid binary ionic mixture in the central regions of a white dwarf reaches the {\it miscibility gap}, an unstable domain where a mechanical mixture of two solids is thermodynamically preferable. As soon as the temperature drops below the {\it solvus}, the original solid
begins nucleating out a new solid solution, whose composition is determined by the phase diagram. As evolution proceeds, the parent mixture continues to decompose until both solutions reach pure compositions. This process is known as {\it exsolution} and, depending on the initial composition of the solid, it can release or absorb heat thus leading to a deceleration or an acceleration of white dwarf cooling. In this paper, we have studied the exsolution process occurring in white dwarf stars with CO and ONe cores, H-rich envelopes, and masses ranging from 0.53 to 1.29 $M_\odot$. Although we have neglected $^{22}$Ne sedimentation, which should lead to further delays in the cooling times, we expect this process to be independent of the exsolution and to not affect our main findings.

We have found that, similar to crystallization, exsolution occurs at brighter luminosities in more massive white dwarfs due to higher mass densities in their interior. On the other hand, in contrast with crystallization, exsolution occurs at brighter luminosities for CO white dwarfs than for ONe white dwarfs of the same mass. The reason for this can be deduced from the phase diagrams for the exsolution \citep[][]{2023MNRAS.522L..26B}, which predict a higher exsolution temperature relative to the melting temperature in CO mixtures. In our CO models, exsolution starts in the center of the star while the outer layers are still crystallizing. 
We have also found that exsolution is slow and produces (or absorbs) noticeable amounts of energy for several Gyrs. It can take $\sim 2.5$ Gyrs for a stellar layer to complete the exsolution process, in contrast with crystallization which we assume to be instantaneous in each layer.

The impact of the exsolution process on the white dwarf cooling times depends on the mass and the exact chemical profile of the solid mixture. Indeed, those layers, where the numerical abundance of the heavier component is larger than the critical value, will absorb heat, while layers, where the numerical abundance of the heavier component is below this number, will release it. We have found that some inner regions of 0.53, 0.58, 0.66, 1.1, 1.16, 1.23, and 1.29 $M_\sun$ CO-core models absorb heat. However, this is counterbalanced with the energy release in the outer layers of these models.

Overall, we can characterize the effect of the exsolution process on the white dwarf cooling times as marginal, with a maximum delay of $\sim 0.35$ Gyrs for luminosities down to $\log(L/L_\odot)\sim -5$, standard compositions, and preferred assumptions regarding the miscibility gap microphysics. For ONe white dwarfs, the cooling delay caused by the exsolution heat release is of the same order of magnitude as that caused by crystallization, when both latent heat and phase separation are included. For the most massive ONe white dwarfs ($M\gtrsim 1.2 M_\odot$), the change in stellar radius induced by phase separation produces a ``reversal effect'' that practically eliminates the cooling delay associated with crystallization. In this case, the exsolution cooling delay becomes the dominant effect.

We have also plotted the location of exsolution and crystallization processes on the {\it Gaia} color magnitude diagram, relying on the atmosphere models for pure H composition of \cite{2010MmSAI..81..921K}. We have not found any branch on this diagram associated with the exsolution, which is consistent with the marginal importance of this process for the white dwarf cooling times. Also, we have found that the so-called Q branch is consistent with the crystallization for both CO and ONe white dwarfs, when pure H-atmospheric composition is considered.

We have compared our models with the IR-faint white dwarfs from \cite{2022ApJ...934...36B}, finding that these objects with masses $M_{\rm WD}/M_\odot \gtrsim  0.66$ should be undergoing exsolution. We have found that only 7 of the 105 IR-faint white dwarfs in this sample are in the Debye cooling regime, where the heat capacity drops due to ion quantum effects and cooling accelerates.

{  The time delay induced by exsolution in massive white dwarfs 
might have significant implications for white dwarf cosmochronology. 
Within a population of white dwarfs, the less massive ones have had less 
time to evolve to very low luminosities compared to their more massive counterparts. 
Consequently, the faint end of the white dwarf luminosity function is populated 
by massive white dwarfs, where exsolution leads to an approximately 0.3 Gyr delay. 
Therefore, the age estimated by the cutoff of the white dwarf luminosity function 
in such populations could have been underestimated by approximately 0.3 Gyrs.}

In this preliminary work, we have not considered a chemical redistribution caused by exsolution, which could also lead to a further delay on the white dwarf cooling times. The chemical redistribution is known to accompany exsolution in binary solid mixtures on Earth. However, given the solid state of the white dwarf matter, this effect is hard to model reliably and it may be postponed  till really low luminosities.   

Summarizing, it would be very hard to prove the exsolution occurrence in white dwarfs with current observational techniques but the situation may improve in the future. 


\begin{acknowledgements}

{  We acknowledge the expert referee, Maurizio Salaris, whose comments helped to improve the original version of the manuscript.}
MC acknowledges
grant RYC2021-032721-I, funded by MCIN/AEI/10.13039/501100011033 and by the European Union NextGenerationEU/PRTR.
  This work was partially supported by the AGAUR/Generalitat de Catalunya grant SGR-386/2021 and by the Spanish MINECO grant PID2020-117252GB-I00.
 This  research  made use of  NASA Astrophysics Data System and data from the European Space Agency (ESA) mission {\it Gaia} (\url{https://www.cosmos.esa.int/gaia}), processed by the {\it Gaia} Data Processing and Analysis Consortium (DPAC, \url{https://www.cosmos.esa.int/web/gaia/dpac/consortium}). Funding for the DPAC has been provided by national institutions, in particular the institutions participating in the {\it Gaia} Multilateral Agreement. 
 
\end{acknowledgements}

\bibliographystyle{aa}
\bibliography{ultramassiveCO}



\end{document}